\documentclass[11pt]{article}
\usepackage{moriond,epsfig}

\usepackage{microtype}
\usepackage{xspace} 

\usepackage{graphicx}  
\graphicspath{{./figs/}} 
\usepackage[percent]{overpic}

\usepackage[usenames,dvipsnames]{color}

\usepackage{amsmath} 
\usepackage{amssymb}
\usepackage{amsfonts}
\usepackage{upgreek} 

\usepackage{hyperref}    
\usepackage[all]{hypcap} 

\usepackage{ifthen} 
\newboolean{pdflatex}
\setboolean{pdflatex}{true} 
\newboolean{articletitles}
\setboolean{articletitles}{true} 
\newboolean{uprightparticles}
\setboolean{uprightparticles}{false} 

\def\ux85 {\mbox{UX85}\xspace}

\ifthenelse{\boolean{uprightparticles}}%
{

 \def\Ppi         {\ensuremath{\uppi}\xspace}

 \def\PDelta      {\ensuremath{\Delta}\xspace}                 
 \def\PXi      {\ensuremath{\Xi}\xspace}                 
 \def\PLambda      {\ensuremath{\Lambda}\xspace}                 
 \def\PSigma      {\ensuremath{\Sigma}\xspace}                 
 \def\POmega      {\ensuremath{\Omega}\xspace}                 
 \def\PUpsilon      {\ensuremath{\Upsilon}\xspace}                 
 
 \def\PB      {\ensuremath{\mathrm{B}}\xspace}                 
                  
 \def\PD      {\ensuremath{\mathrm{D}}\xspace}

 \def\PK      {\ensuremath{\mathrm{K}}\xspace}

 \def\Pi      {\ensuremath{\mathrm{i}}\xspace}

}
{

 \def\Ppi         {\ensuremath{\pi}\xspace}

 \mathchardef\PDelta="7101
 \mathchardef\PXi="7104
 \mathchardef\PLambda="7103
 \mathchardef\PSigma="7106
 \mathchardef\POmega="710A
 \mathchardef\PUpsilon="7107
                  
 \def\PB      {\ensuremath{B}\xspace}                 
                  
 \def\PD      {\ensuremath{D}\xspace}

 \def\PK      {\ensuremath{K}\xspace}

 \def\Pi      {\ensuremath{i}\xspace}

}

\def\g      {\ensuremath{\Pgamma}\xspace}

\def\pion  {\ensuremath{\Ppi}\xspace}

\def\pip   {\ensuremath{\pion^+}\xspace}
\def\pim   {\ensuremath{\pion^-}\xspace}

\def\kaon  {\ensuremath{\PK}\xspace}
  \def\Kbar  {\kern 0.2em\overline{\kern -0.2em \PK}{}\xspace}

\def\Kz    {\ensuremath{\kaon^0}\xspace}
\def\Kzb   {\ensuremath{\Kbar^0}\xspace}
\def\KzKzb {\ensuremath{\Kz \kern -0.16em \Kzb}\xspace}
\def\Kp    {\ensuremath{\kaon^+}\xspace}
\def\Km    {\ensuremath{\kaon^-}\xspace}

\def\KpKm  {\ensuremath{\Kp \kern -0.16em \Km}\xspace}
\def\KS    {\ensuremath{\kaon^0_{\rm\scriptscriptstyle S}}\xspace}

  \def\Dbar    {\kern 0.2em\overline{\kern -0.2em \PD}{}\xspace}
\def\D       {\ensuremath{\PD}\xspace}

\def\Dz      {\ensuremath{\D^0}\xspace}
\def\Dzb     {\ensuremath{\Dbar^0}\xspace}
\def\DzDzb   {\ensuremath{\Dz {\kern -0.16em \Dzb}}\xspace}
\def\Dp      {\ensuremath{\D^+}\xspace}
\def\Dm      {\ensuremath{\D^-}\xspace}

\def\DpDm    {\ensuremath{\Dp {\kern -0.16em \Dm}}\xspace}

\def\B       {\ensuremath{\PB}\xspace}
\def\Bbar    {\ensuremath{\kern 0.18em\overline{\kern -0.18em \PB}{}}\xspace}

\def\Bu      {\ensuremath{\B^+}\xspace}
\def\Bub     {\ensuremath{\B^-}\xspace}
\def\Bp      {\ensuremath{\Bu}\xspace}
\def\Bm      {\ensuremath{\Bub}\xspace}
\def\Bpm     {\ensuremath{\B^\pm}\xspace}

  \def\Y#1S{\ensuremath{\PUpsilon{(#1S)}}\xspace}

\def\Lbar {\ensuremath{\kern 0.1em\overline{\kern -0.1em\PLambda}}\xspace}

\newcommand{\decay}[2]{\ensuremath{#1\!\to #2}\xspace}         

\def\to                 {\ensuremath{\rightarrow}\xspace}

\def\CP                {\ensuremath{C\!P}\xspace}

\def\AT#1     {\ensuremath{A_{\mathrm{T}}^{#1}}\xspace}           

\def\C#1      {\ensuremath{\mathcal{C}_{#1}}\xspace}                       
\def\Cp#1     {\ensuremath{\mathcal{C}_{#1}^{'}}\xspace}                    
\def\Ceff#1   {\ensuremath{\mathcal{C}_{#1}^{\mathrm{(eff)}}}\xspace}        
\def\Cpeff#1  {\ensuremath{\mathcal{C}_{#1}^{'\mathrm{(eff)}}}\xspace}       
\def\Ope#1    {\ensuremath{\mathcal{O}_{#1}}\xspace}                       
\def\Opep#1   {\ensuremath{\mathcal{O}_{#1}^{'}}\xspace}                    

\newcommand{\tev}{\ensuremath{\mathrm{\,Te\kern -0.1em V}}\xspace}
\newcommand{\gev}{\ensuremath{\mathrm{\,Ge\kern -0.1em V}}\xspace}
\newcommand{\mev}{\ensuremath{\mathrm{\,Me\kern -0.1em V}}\xspace}
\newcommand{\kev}{\ensuremath{\mathrm{\,ke\kern -0.1em V}}\xspace}
\newcommand{\ev}{\ensuremath{\mathrm{\,e\kern -0.1em V}}\xspace}
\newcommand{\gevc}{\ensuremath{{\mathrm{\,Ge\kern -0.1em V\!/}c}}\xspace}
\newcommand{\mevc}{\ensuremath{{\mathrm{\,Me\kern -0.1em V\!/}c}}\xspace}
\newcommand{\gevcc}{\ensuremath{{\mathrm{\,Ge\kern -0.1em V\!/}c^2}}\xspace}
\newcommand{\gevgevcccc}{\ensuremath{{\mathrm{\,Ge\kern -0.1em V^2\!/}c^4}}\xspace}
\newcommand{\mevcc}{\ensuremath{{\mathrm{\,Me\kern -0.1em V\!/}c^2}}\xspace}

\def\invfb   {\ensuremath{\mbox{\,fb}^{-1}}\xspace}

\def\gsim{{~\raise.15em\hbox{$>$}\kern-.85em
          \lower.35em\hbox{$\sim$}~}\xspace}
\def\lsim{{~\raise.15em\hbox{$<$}\kern-.85em
          \lower.35em\hbox{$\sim$}~}\xspace}

\def\tell1  {TELL1\xspace}
\def\ukl1   {UKL1\xspace}

\usepackage{cite} 
\usepackage{mciteplus}

\newcommand{\figs}{figs}

\newcommand{\omcl}{\ensuremath{1-{\rm CL}}\xspace}

\newcommand{\gDKCentral}	{\ensuremath{72.0}\xspace}

\newcommand{\gDKOnesig} 	{\ensuremath{[   56.4,    86.7]}\xspace}
\newcommand{\gDKTwosig} 	{\ensuremath{[   42.6,    99.6]}\xspace}

\newcommand{\gDpiCentralI}  {\ensuremath{18.9}\xspace}

\newcommand{\gDpiOnesigIC}	 {\ensuremath{[7.4, 99.2]}\xspace} 
\newcommand{\gDpiOnesigIIC}	 {\ensuremath{[167.9, 176.4]}\xspace}

\newcommand{\gDKDpiCentral}    {\ensuremath{72.6}\xspace}

\newcommand{\gDKDpiOnesigC}		{\ensuremath{[55.4, 82.3]}\xspace} 
\newcommand{\gDKDpiTwosigC}	 	{\ensuremath{[40.2, 92.7]}\xspace}

\newcommand{\BpmDh}   {\texorpdfstring{\decay{\Bpm}{D h^\pm}}			{B+- -> Dh+-}}
\newcommand{\BpmDpi}  {\texorpdfstring{\decay{\Bpm}{D \pi^\pm}}		{B+- -> Dpi+-}}
\newcommand{\BpmDK}   {\texorpdfstring{\decay{\Bpm}{D K^\pm}}			{B+- -> DK+-}}

\newcommand{\DKpi}     {\texorpdfstring{\ensuremath{D\to K\pi}}{D -> Kpi}\xspace}

\newcommand{\DzKShh}   {\texorpdfstring{\ensuremath{\Dz\to \KS h^\pm h^\mp}}{D0 -> KShh}\xspace}

\newcommand{\Dzhh}     {\texorpdfstring{\ensuremath{\Dz\to h^+h^-}}{D -> hh}\xspace}

\newcommand{\DzK}      {\texorpdfstring{\ensuremath{DK^\pm}}{DK}\xspace}
\newcommand{\Dzpi}     {\texorpdfstring{\ensuremath{D\pi^\pm}}{Dpi}\xspace}

\newcommand{\Kpipipi}	{\texorpdfstring{\ensuremath{K^\pm\pi^\mp\pip\pim}}{Kpipipi}\xspace}

\renewcommand{\g}{\texorpdfstring{\ensuremath{\gamma}}{gamma}\xspace}

\newcommand{\rbh}  {\texorpdfstring{\ensuremath{r_B^h}\xspace}{rBh}}
\newcommand{\rbhsq}{\texorpdfstring{\ensuremath{(r_B^{h})^2}\xspace}{rBh2}}
\newcommand{\dbh}  {\ensuremath{\delta_B^h}\xspace}

\newcommand{\rb}  {\texorpdfstring{\ensuremath{r_B^K}\xspace}{rBK}}

\newcommand{\db}  {\ensuremath{\delta_B^K}\xspace}

\newcommand{\rbpi}  {\texorpdfstring{\ensuremath{r_B^{\pi}}\xspace}{rBpi}}

\newcommand{\dbpi}  {\ensuremath{\delta_B^{\pi}}\xspace}

\newcommand{\rdFlv}  {\ensuremath{r_{f}}\xspace}
\newcommand{\rdFlvsq}{\ensuremath{r_{f}^2}\xspace}
\newcommand{\ddFlv}  {\ensuremath{\delta_{f}}\xspace}

\newcommand{\ddKpi}  {\ensuremath{\delta_{K\pi}}\xspace}

\newcommand{\xd}{\texorpdfstring{\ensuremath{x_D}\xspace}{xD}}
\newcommand{\yd}{\texorpdfstring{\ensuremath{y_D}\xspace}{yD}}

\newcommand{\Rads}{\ensuremath{R_{\rm ADS}}\xspace}
\newcommand{\Aads}{\ensuremath{A_{\rm ADS}}\xspace}

\begin{document}


\title{Measurement of \g from \BpmDh decays at LHCb---including the effect of \Dz--\Dzb mixing}
\author{T.M. Karbach \footnote{on behalf of the LHCb collaboration}}
\address{CERN}
\maketitle

\begin{quote}
\small I present a measurement of the CKM angle \g from a combination
of three LHCb measurements using the tree decays \BpmDK and \BpmDpi.
These measurements are based on a dataset corresponding to $1.0\invfb$,
collected in 2011. In contrast to what was presented at FPCP,
these proceedings fully include the effect of \Dz--\Dzb mixing on
the combination.
I also report on the inclusion of new, preliminary results on the Dalitz analysis 
of \BpmDK, \DzKShh, using a dataset corresponding to $3\invfb$, collected
in 2011--2012. 
\end{quote}

\begin{quote}
\begin{center}
\small {\it presented at}\\
FPCP2013, Flavor Physics \& \CP Violation, Buzios, Rio de Janeiro, Brazil,\\
May 19$^{th}$--24$^{th}$, 2013.
\end{center}
\end{quote}


\section{Introduction}

The CKM angle \g is the least well measured angle of the unitarity triangle
of the CKM matrix. The existing
constraints come mainly from the $B$ factories BaBar and Belle,
and mainly use analyses of tree-level \BpmDK decays. Both collaborations
have independently combined all their relevant measurements in a frequentist
way, and arrive at $\g = (69^{+17}_{-16})^\circ$ (BaBar~\cite{Lees:2013fk}) 
and $\g = (68^{+15}_{-14})^\circ$ (Belle~\cite{Trabelsi:2013uj}).

The LHCb collaboration now has results on \g available as well. 
As I presented at FPCP, and summarize in these proceedings, these
too are combined in a frequentist way, and already with $1\invfb$ of 2011 data LHCb reaches
a precision comparable to that of the $B$ factories. For this not only
the traditional \BpmDK decays are used, but also \BpmDpi decays.
Up to now these were mostly seen as pure normalization channels as their
sensitivity to \g is smaller than that of \BpmDK decays. But they do
contribute to the overall precision.

Although \g is still statistically limited, its uncertainty is steadily
decreasing as new measurements appear. At some point systematic effects,
that traditionally were neglected, need
to be explicitly considered. Most prominently this is the effect
of \Dz--\Dzb mixing, which already now affects the analysis of 
\BpmDpi decays. But also the possible effect of \CP violation in
\Dz decays plays a role. As a result of fruitful discussions held at FPCP, 
the combined \g measurement by LHCb is now the first combination of measurements that 
explicitly considers both effects.

Anticipating the following sections, LHCb measures a value of
$\g=(72.6^{+9.1}_{-15.9})^\circ$~\cite{Aaij:2013zfa}. The agreement between the combinations
by Belle, BaBar, and LHCb is impressive. It is very reassuring to see the theoretical framework
under good control, and to see the efforts of over a decade of $e^+e^-$ physics
and results obtained in LHC's hadronic environment coming together so nicely. 
It is important, however, to keep in mind,
that the three values are not entirely independent---they all share
input measurements on the hadronic \Dz parameters. One prominent such parameter is the
strong phase difference \ddKpi between the decay amplitudes of
$\Dz\to\Km\pip$ and $\Dzb\to\Km\pip$, which is, in the important ADS
measurement of \BpmDK, \DKpi, additive to \g but also not very precisely known.
For example, the latest HFAG \CP-violation-allowed average is, in our phase convention,
$\ddKpi = (199.5^{+8.6}_{-11.1})^\circ$~\cite{HFAG}.

\section{Input measurements using $1.0\invfb$}

\newcommand{\fbar}{\ensuremath{\overline{f}}\xspace}

LHCb has published three measurements of \BpmDK decays relevant to this \g measurement.
The first two are a GLW/ADS measurement using two-body \Dz final states~\cite{Aaij:2012kz},
and an ADS measurement using the four-body \Dz final state \Kpipipi~\cite{Aaij:2013aa}.
Both of these measurements also use \BpmDpi decays, in addition to the 
established \BpmDK decays. The third measurement is
a Dalitz-model independent GGSZ analysis of \DzKShh decays~\cite{Aaij:2012hu}.
All three measurements are based on $1\invfb$ of data recorded in 2011,
constituting about one third of the data recorded to date.

The above three main inputs are combined with information on the hadronic
\Dz parameters, on \CP violation in \Dzhh decays, and on the \Dz--\Dzb mixing 
parameters, taken from CLEO~\cite{Lowery:2009id}, HFAG~\cite{HFAG}, 
and LHCb~\cite{LHCb_Dmixing}, respectively.

\section{Effect of \Dz--\Dzb mixing}

It was known since many years that \Dz--\Dzb mixing complicates the measurement
of \g from \BpmDK decays (and \BpmDpi decays)~\cite{Silva:1999bd,Atwood:2000ck,Grossman:2005rp,Bondar:2010qs}.
Traditionally, the effect
was neglected, as it was estimated to be small compared to the achievable uncertainty
on \g. But it was clear that, in order to reach a degree-precision on \g, the
effect of \Dz--\Dzb mixing cannot be neglected anymore. It is
straightforward to account for mixing in the relevant equations,
fully retaining the outstanding theoretical cleanliness of the \g measurement.

Recently, the \Dz--\Dzb mixing effects on \g measurements were nicely
summarized in Ref.~\cite{Matteo}. In general, the effect is larger for \BpmDpi
decays than for \BpmDK decays, because the amplitude ratio \rbpi is about
an order of magnitude smaller than \rb. At leading order in the mixing parameters $x$ and
$y$, and neglecting \CP violation in mixing, (i) the GLW measurements are unaffected for
both \DzK and \Dzpi,
(ii) the \DzK model-independent GGSZ measurement is unaffected, (iii) the
\DzK ADS analysis is affected at the degree level, and (iv) the
\Dzpi ADS analysis is largely affected, as the terms describing \Dz--\Dzb
mixing are of the same order of magnitude as the interference terms
giving the sensitivity to \g.

The \Dz--\Dzb mixing effects also depend on the selection criteria that
were used to select \BpmDh decays~\cite{Matteo}. Since the \Dz decay time is a powerful
discriminating variable, the \Dz decay time acceptance will not be flat,
which needs to be considered in the measurement of \g.

The ADS observables which receive mixing corrections are the charge-averaged 
ratios of \BpmDK and \BpmDpi decays
\begin{align}
R_{K/\pi}^{f} &= \frac{\Gamma(\Bm\to D[\to f]K^-)   + \Gamma(\Bp\to D[\to \fbar]K^+)}
  	                  {\Gamma(\Bm\to D[\to f]\pi^-) + \Gamma(\Bp\to D[\to \fbar]\pi^+)}\,, \label{eq:rkpi}
\end{align}
where $f$ is the relevant final state. The ratios $R_{K/\pi}^{f}$
are related to \g and the hadronic parameters through
\begin{align}
R_{K/\pi}^{f} &= R_{\rm cab} 
                 \frac{1 + (\rb  \rdFlv)^2 + 2 \rb   \rdFlv \kappa\cos(\db   - \ddFlv) \cos\gamma + M^{K}_- + M^{K}_+}
                      {1 + (\rbpi\rdFlv)^2 + 2 \rbpi \rdFlv \kappa\cos(\dbpi - \ddFlv) \cos\gamma + M^{\pi}_- + M^{\pi}_+}\,, \label{eq:rkpi1}
\end{align}
where $\kappa$ denotes the coherence factor.
The $D$ mixing correction terms $M^h_{\pm}$ are~\cite{Silva:1999bd}
\begin{align}
 M^h_{\pm} &= \left(\kappa\rdFlv(\rbhsq-1)\sin\ddFlv + \rbh(1-\rdFlvsq)\sin(\dbh\pm\g)\right)\, a_D \, \xd\nonumber\\
           &- \left(\kappa\rdFlv(\rbhsq+1)\cos\ddFlv + \rbh(1+\rdFlvsq)\cos(\dbh\pm\g)\right)\, a_D \, \yd\,.
\end{align}
The coefficient $a_D$ parameterizes the effect of a finite \Dz decay 
resolution and non-flat acceptance. It takes the value 
of $a_D=1$ in case of an ideal,
flat acceptance and negligible time resolution. For a 
realistic acceptance and resolution model
present in the two- and four-body GLW/ADS analyses of Refs.~\cite{Aaij:2012kz,Aaij:2013aa}, 
it is estimated to be $a_D=1.20\pm0.04$.
The charge asymmetries are
\begin{align}
	A_{h}^{f} &= \frac{\Gamma(\Bm\to D[\to f]h^-) - \Gamma(\Bp\to D[\to \fbar]h^+)}
	                  {\Gamma(\Bm\to D[\to f]h^-) + \Gamma(\Bp\to D[\to \fbar]h^+)} \,, 
\end{align}
which are related to \g and the hadronic parameters through
\begin{align}
    A_{h}^{f} &= \frac{2 \rbh \rdFlv \kappa\sin(\dbh - \ddFlv) \sin\gamma + M^{h}_- - M^{h}_+}
                      {1 + (\rbh\rdFlv)^2 + 2 \rbh \rdFlv \kappa\cos(\dbh - \ddFlv) \cos\gamma + M^{h}_- + M^{h}_+}\,, \label{eq:acp1}
\end{align}
where \rbh denotes \rb and \rbpi.
Finally, the non charge-averaged ratios of suppressed and 
favored $D$ final states are
\begin{align}
R_{h}^{\pm}  &= \frac{\Gamma(\Bpm\to D[\to f_{\rm sup}]h^\pm)}
	                 {\Gamma(\Bpm\to D[\to f]h^\pm)} \nonumber\\
	         &= \frac{\rdFlvsq + \rbhsq + 2 \rbh \rdFlv\kappa\cos(\dbh+\ddFlv\pm\gamma) - [M^{h}_{\pm}]_{\rm sup}}
	                 {1 + (\rbh\rdFlv)^2 + 2 \rbh \rdFlv\kappa\cos(\dbh-\ddFlv\pm\gamma) + M^{h}_{\pm}}\,,\label{eq:rpm}
\end{align}
where $f_{\rm sup}$ is the suppressed final state,
and $f$ the allowed one.
The suppressed $D$ mixing correction terms are given,
at leading order in \xd and \yd, by
\begin{align}
  [M^{h}_{\pm}]_{\rm sup}&= \left(\kappa\rdFlv(\rbhsq-1)\sin\ddFlv + \rbh(1-\rdFlvsq)\sin(\dbh\pm\g)\right)\, a_D \, \xd\nonumber\\
                         & +\left(\kappa\rdFlv(\rbhsq+1)\cos\ddFlv + \rbh(1+\rdFlvsq)\cos(\dbh\pm\g)\right)\, a_D \, \yd\,.
\end{align}

\section{Results on \BpmDK and \BpmDpi, using $1.0\invfb$}

\begin{figure}[!htb]
\centering
\begin{overpic}[width=.35\textwidth]{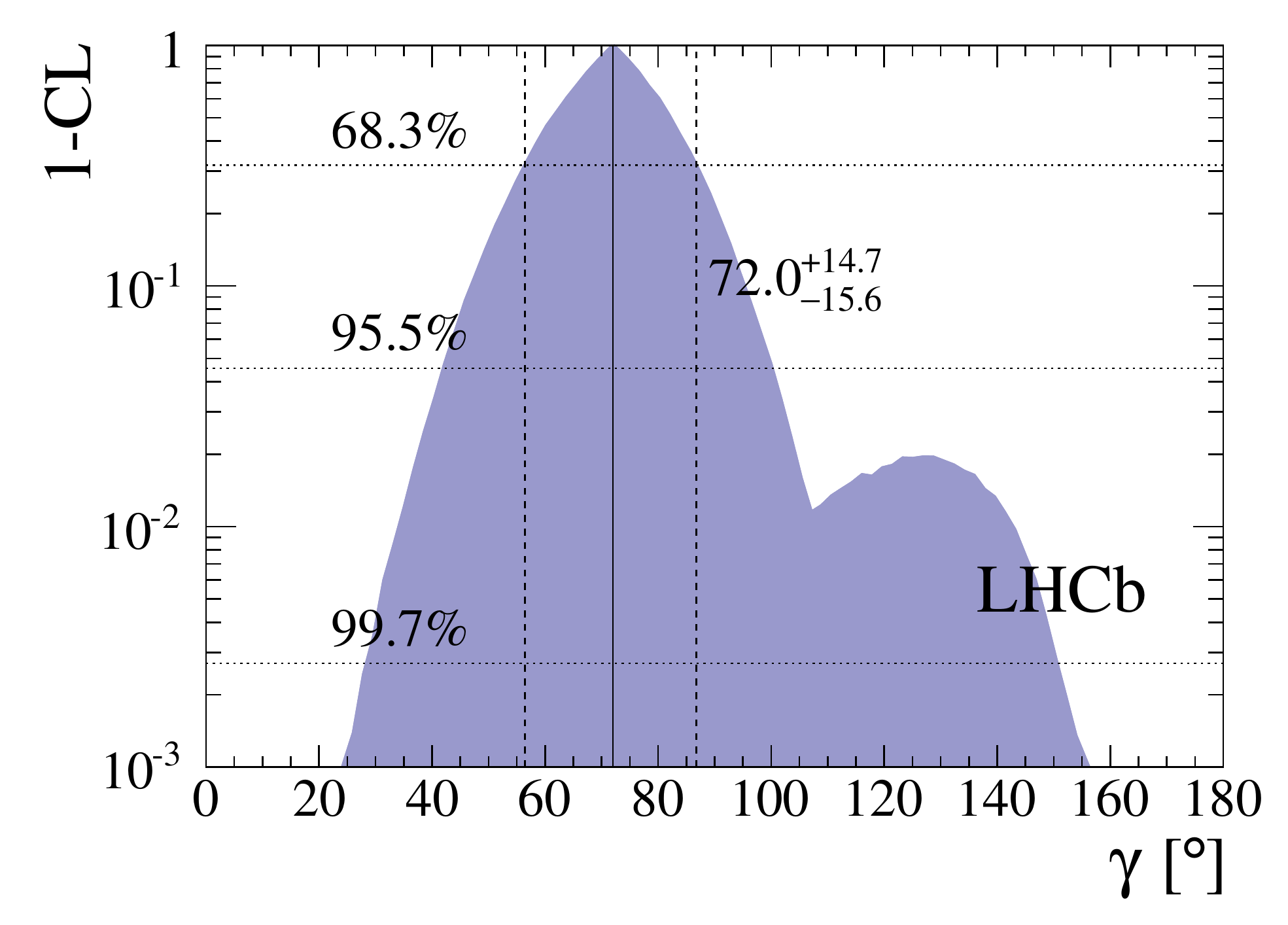}  \put(85,57){a)}\end{overpic}
\begin{overpic}[width=.35\textwidth]{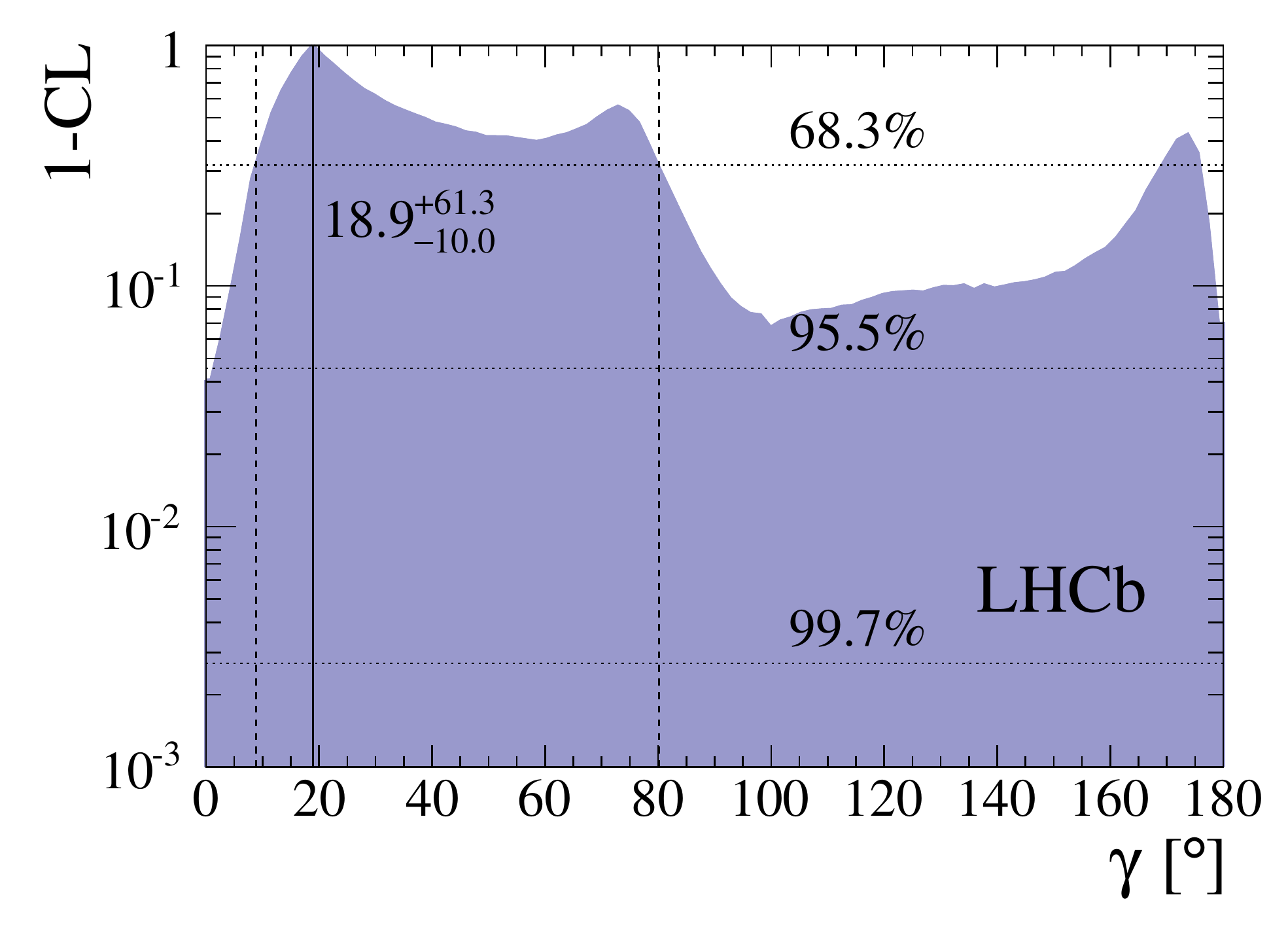} \put(85,57){b)}\end{overpic}
\begin{overpic}[width=.35\textwidth]{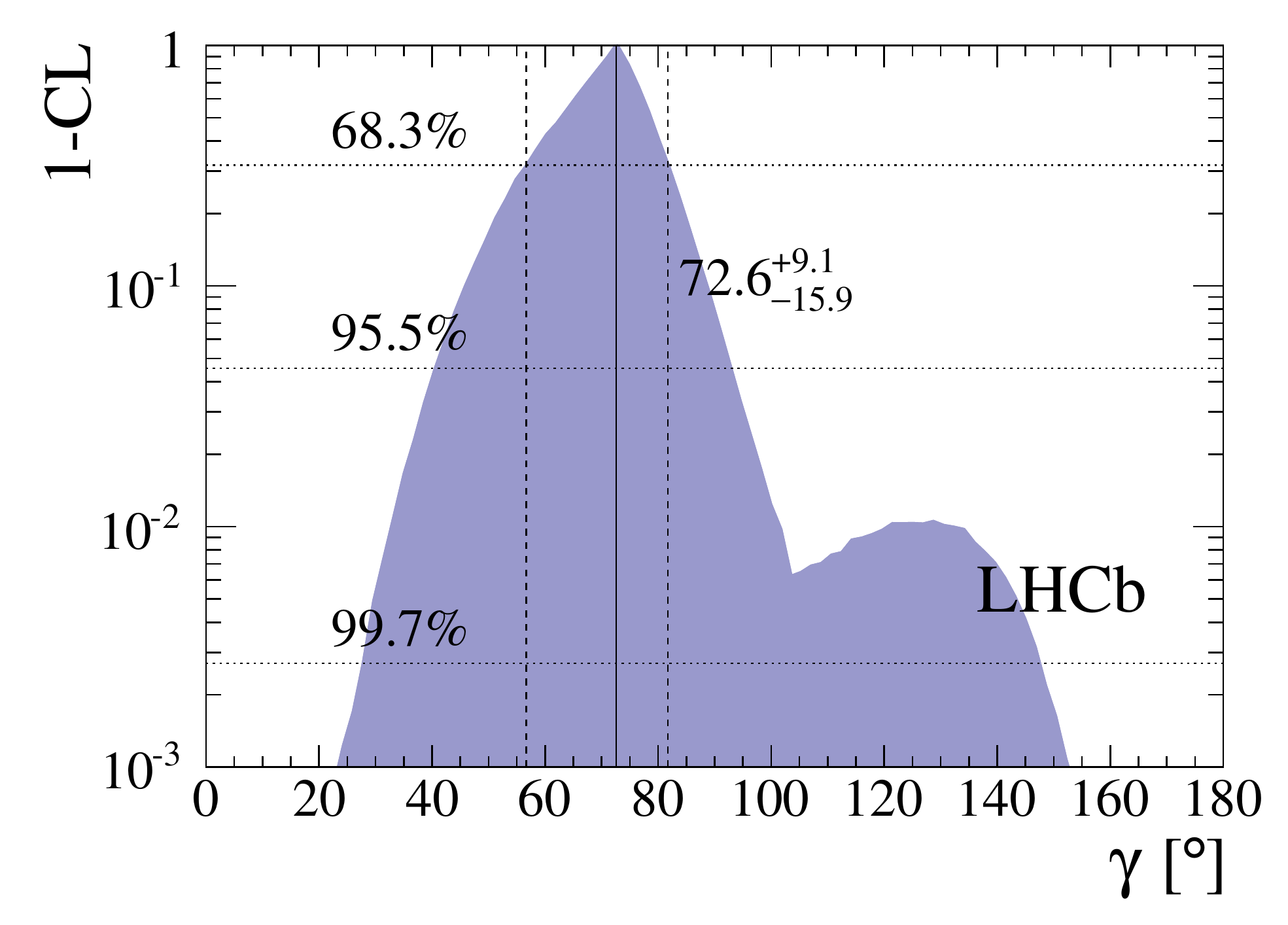}   \put(85,57){c)}\end{overpic}
\caption{\small Graphs showing \omcl for \g, for (a) the \DzK 
combination, for (b) the \Dzpi combination, and (c) for the full
\DzK and \Dzpi combination.
The reported numbers correspond to the best-fit values and the uncertainties
are computed using the respective $68.3\%$ CL confidence intervals (not corrected
for undercoverage and neglected systematic correlations).}
\label{fig:dk1log}
\end{figure}

\begin{figure}[!htb]
\centering
\begin{overpic}[width=.35\textwidth]{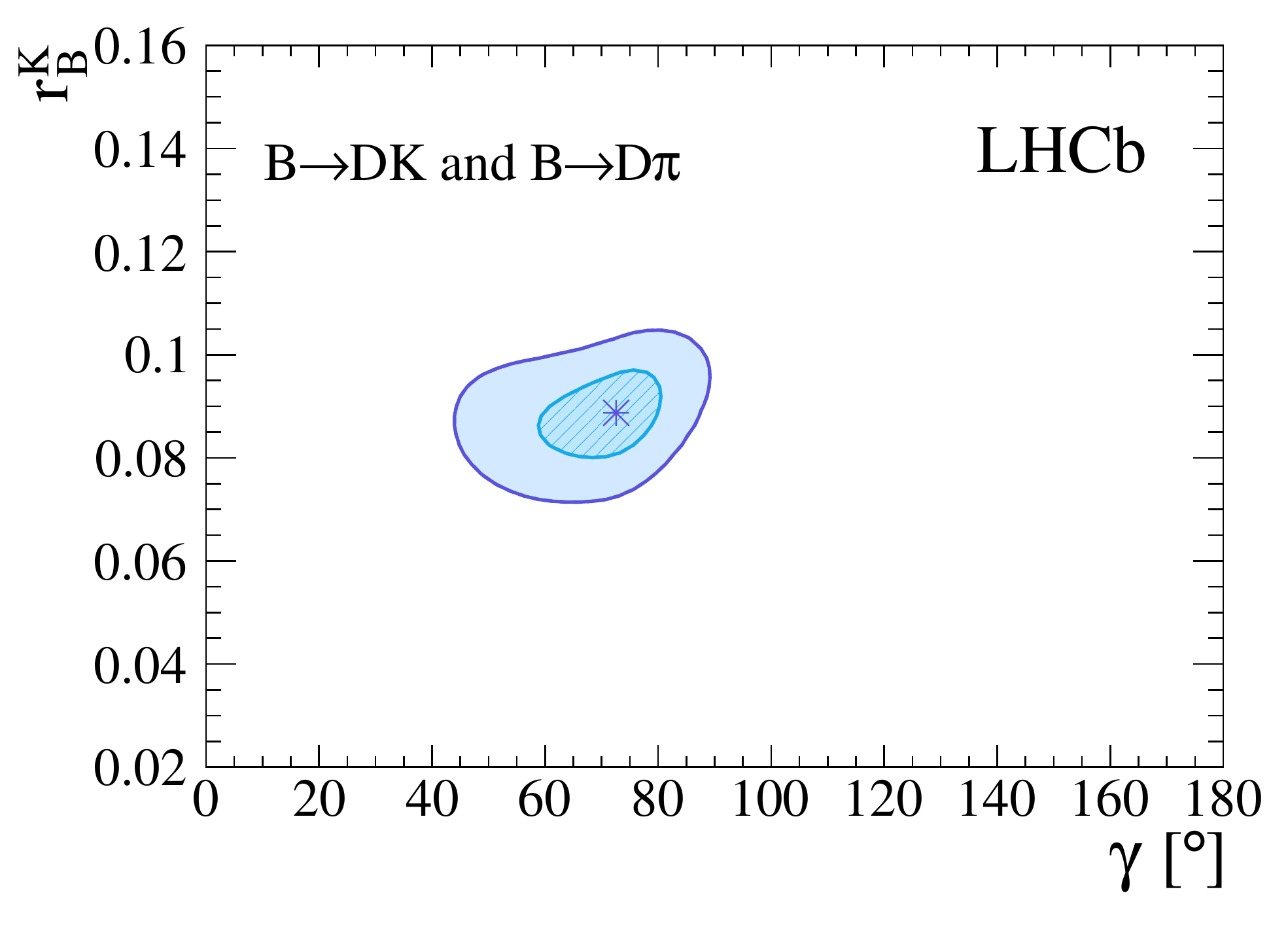} \put(85,47){a)}\end{overpic}
\begin{overpic}[width=.35\textwidth]{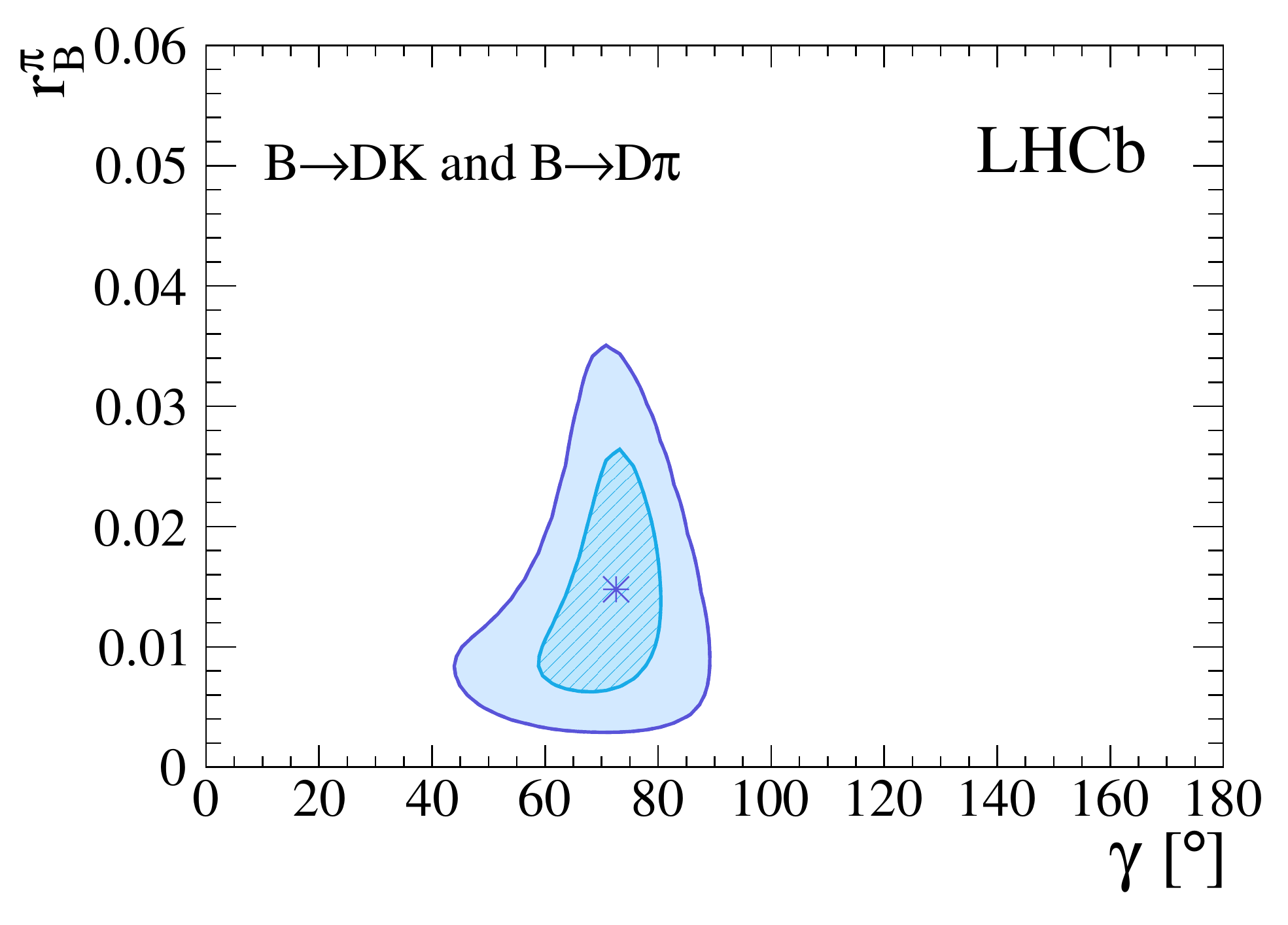}\put(85,47){b)}\end{overpic}
\caption{\small Profile likelihood contours of 
(a) \g vs.~\rb, and
(b) \g vs.~\rbpi, 
for the full
\DzK and \Dzpi combination.
The contours are the $n\sigma$ 
profile likelihood contours, where $\Delta\chi^2=n^2$ with $n=1,2$.
The markers denote the best-fit values.
}
\label{fig:2dplots1}
\end{figure}

The combination follows a frequentist procedure (plug-in method)
and is fully described in Ref.~\cite{Aaij:2013zfa}. The frequentist coverage is 
computed at the best-fit point, and the final confidence intervals
are enlarged to correct for a small undercoverage. The results for \g
are shown in Table~\ref{tab:results} for three cases: only combining \BpmDK inputs,
only combining \BpmDpi inputs, and the full \DzK and \Dzpi combination.
The \BpmDpi decays do indeed add sensitivity to the full combination.
This can be explained by the observed high value of \rbpi, that drives
the sensitivity, $\rbpi = 0.015^{+0.012}_{-0.009}$
at 68\% CL (obtained from the full combination), which is much larger 
than (but consistent with) the expectation
of $\rbpi = 0.006$, as estimated from Ref.~\cite{Fleischer:2011uq}.
Figures~\ref{fig:dk1log} and~\ref{fig:2dplots1} illustrate these results.

\begin{table}[!h]
\centering
\caption{\small Confidence intervals and best-fit values for \g of three
combinations.}
\label{tab:results}
\begin{tabular}{lccc}
\hline
Quantity & \DzK combination    & \Dzpi combination      & \DzK and \Dzpi combination\\
\hline
\g       & $\gDKCentral^\circ$ & $\gDpiCentralI^\circ$  & $\gDKDpiCentral^\circ$ \\
68\% CL  & $\gDKOnesig^\circ$  & $\gDpiOnesigIC^\circ\,\cup\,\gDpiOnesigIIC^\circ$ & $\gDKDpiOnesigC^\circ$\\
95\% CL  & $\gDKTwosig^\circ$  & no constraint          & $\gDKDpiTwosigC^\circ$\\
\hline
\end{tabular}
\end{table}

\section{General agreement of input measurements}

The agreement of the input measurements is excellent, leading to
fit probabilities well above 50\%. These are estimated using both a
simple $\chi^2$ test, and a more accurate approach with pseudoexperiments,
and given in Table~\ref{tab:gof}.

\begin{figure}[!htb]
\centering
\includegraphics[width=.45\textwidth]{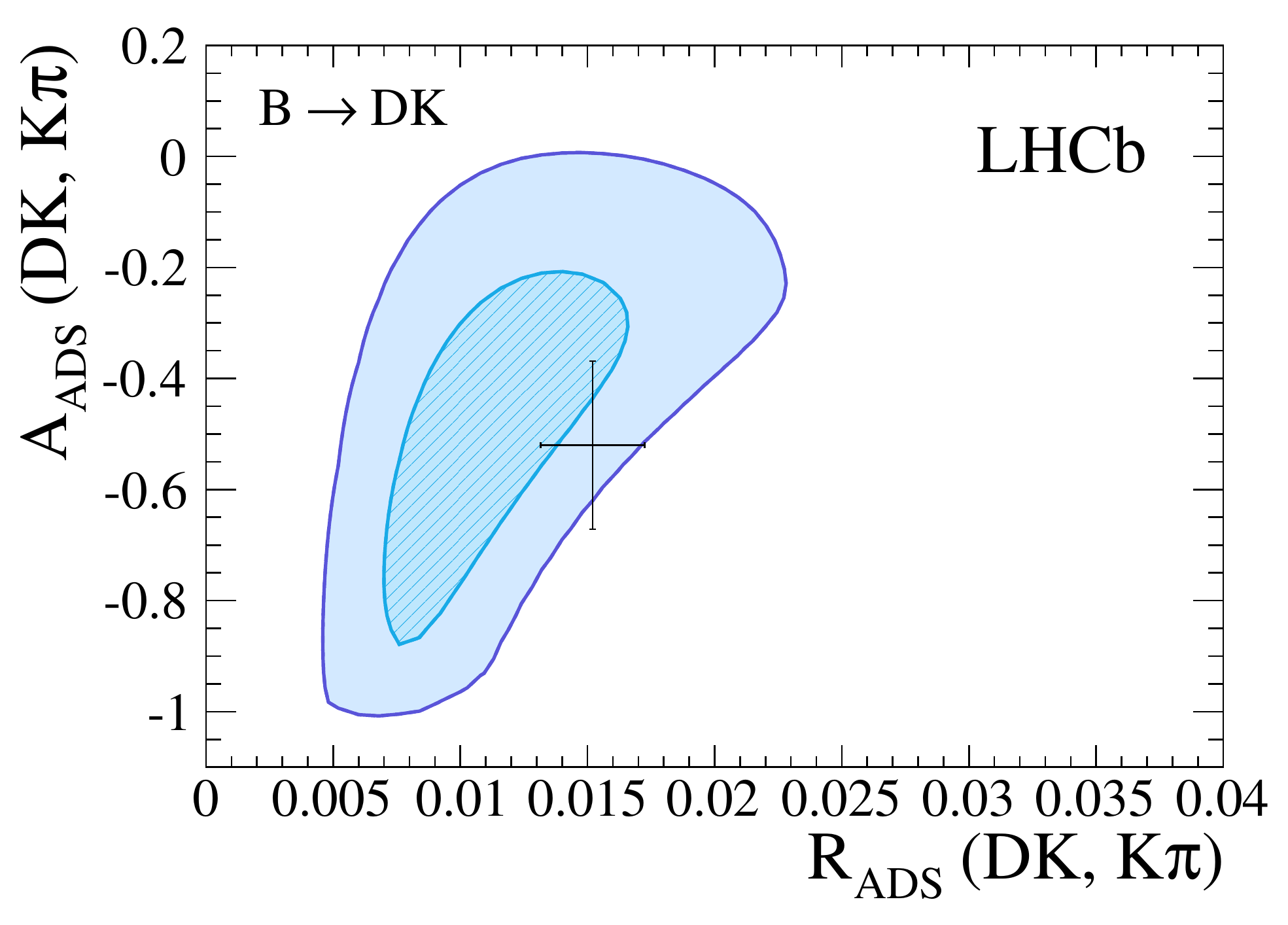}
\includegraphics[width=.35\textwidth]{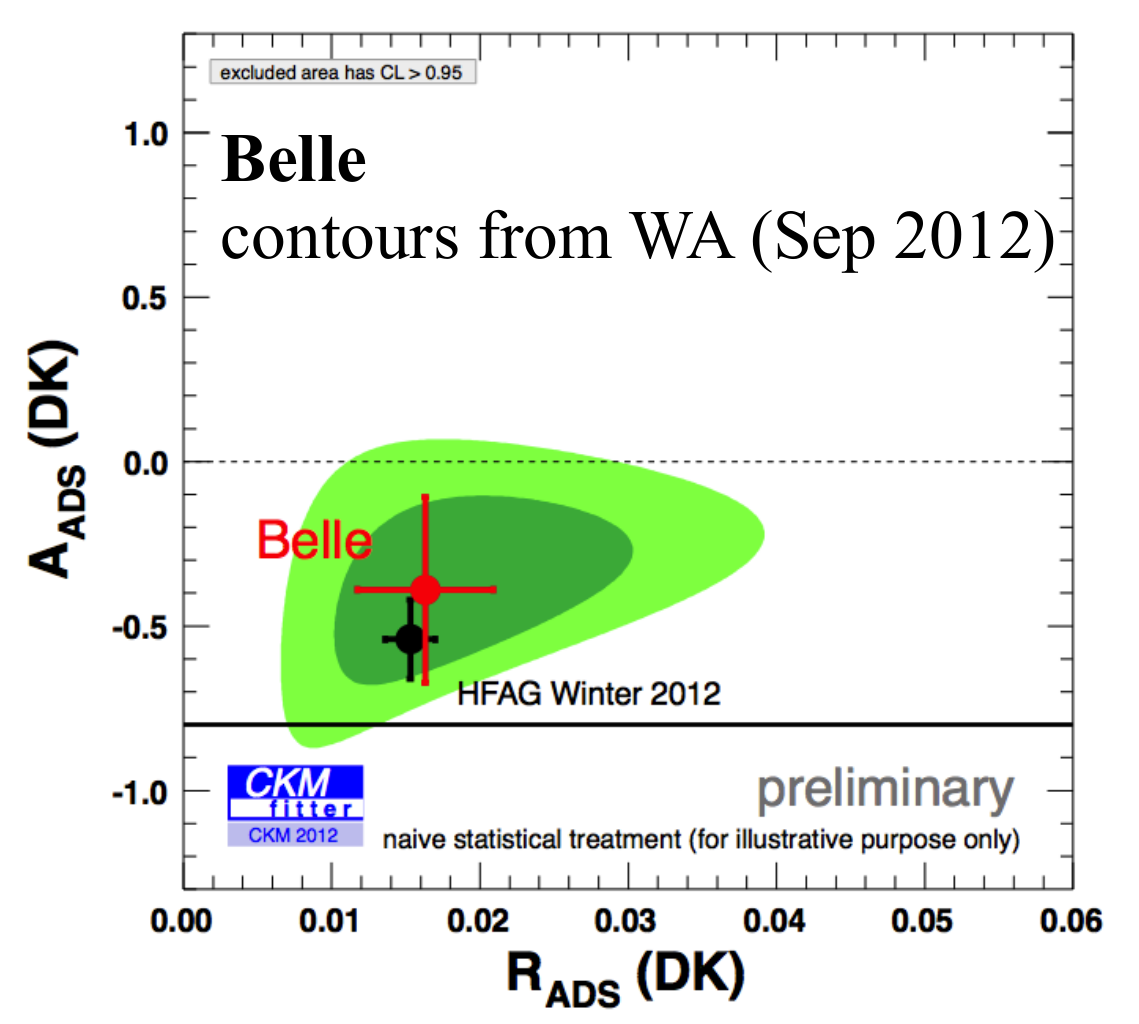}
\caption{\small Prediction of \Rads and \Aads from the full \DzK and
\Dzpi combination, where both \Rads and \Aads were excluded, compared
to their measurements. Left: LHCb~\cite{Aaij:2013aa}, the contours are the $n\sigma$ 
profile likelihood contours, where $\Delta\chi^2=n^2$ with $n=1,2$.
Right: Belle, plot taken from K.~Trabelsi.}
\label{fig:radsaadspred}
\end{figure}

We also made an internal consistency check of the \BpmDK combination, 
inspired by K.~Trabelsi, by excluding the ADS observables \Rads and \Aads
and instead predicting them using all other LHCb inputs.\footnote{Note
that the nominal combination does not use \Rads and \Aads directly, but they
were re-introduced for the purpose of this test.} The prediction agrees
beautifully with the measurements, as shown in Figure~\ref{fig:radsaadspred}.
In fact, Belle sees a very similar picture (also in Fig.~\ref{fig:radsaadspred}).

\begin{table}[!b]
\centering
\caption{\small Numbers of observables $n_{\rm obs}$, numbers of free
parameters in the fit $n_{\rm fit}$, the minimum $\chi^2$ at the
best-fit point, and fit probabilities of the best-fit point for the
three combinations. The quoted uncertainties are due to the limited
number of pseudoexperiments.}
\label{tab:gof}
\begin{tabular}{lccccc}
\hline\\[-2.5ex]
Combination & $n_{\rm obs}$ & $n_{\rm fit}$ & $\chi^2_{\rm min}$ 
& $P$[\%] ($\chi^2$ distribution) & $P$[\%] (pseudoexperiments) \\
\hline\\[-2.5ex]
\DzK  & 29 & 15 & $10.48$ & $72.6$ & $73.9\pm0.2$ \\
\Dzpi & 22 & 14 & $6.28$  & $61.6$ & $61.2\pm0.3$ \\
full  & 38 & 17 & $13.06$ & $90.6$ & $90.9\pm0.1$ \\
\hline                                                    
\end{tabular}                                    
\end{table}

\section{Preliminary results on \BpmDK, \DzKShh, using $3.0\invfb$}

\begin{figure}[!htb]
\centering
\includegraphics[width=.45\textwidth]{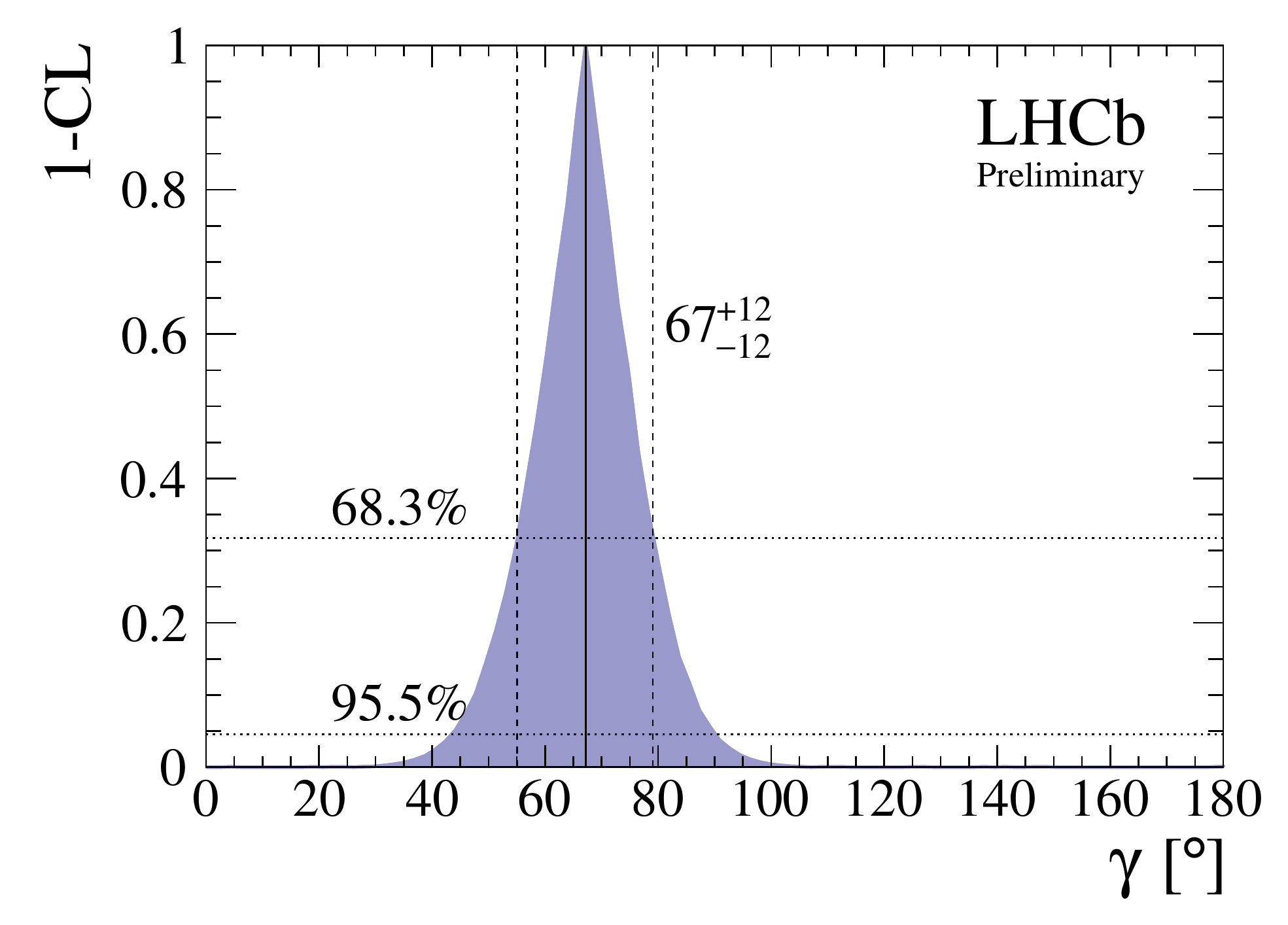}
\caption{\small Graph showing \omcl for \g from \BpmDK decays only, using
the $3\invfb$ GGSZ result~\cite{Aaij:2013qy}, and the $1\invfb$ GLW/ADS results~\cite{Aaij:2012kz,Aaij:2013aa}.}
\label{fig:prelim}
\end{figure}

Going beyond the published analyses using $1\invfb$, LHCb has also a new
preliminary result of the GGSZ analysis of \BpmDK, \DzKShh, using $3.0\invfb$~\cite{Aaij:2013qy}.
When these are used in the \DzK only combination, the best single-experiment
precision for \g is achieved: $\g = (67\pm 12)^\circ$. Figure~\ref{fig:prelim}
shows the corresponding, quite Gaussian \omcl graph.

\section{Conclusion}

The CKM angle \g is now being constrained by three collaborations, Babar, Belle,
and LHCb. It wasn't until recently that all have published their individual averages, 
so that now all three can be seen side by side: They all reach similar precision,
and they all agree. Despite this being a big success, it is clear that one wants
to go beyond the current precision, aiming at a degree uncertainty on \g. Along this
way it is mandatory to correctly consider small systematic effects, that have been
neglected so far---most notably the effect of \Dz--\Dzb mixing. The LHCb combination
is the first to do so.

\section*{References}
\bibliographystyle{LHCb}
\bibliography{references}
\end{document}